\documentclass[dvips]{elsart}
\usepackage{graphics,epsfig}
\newcommand{\thega}{\theta_{\gamma}}
\begin{document}
\begin{frontmatter}
\title{Bound on the tau neutrino magnetic moment from the TRISTAN experiments}
\author[okayama]{N. Tanimoto},
\author[okayama]{I. Nakano} and
\author[sakuda]{M. Sakuda}
\address[okayama]{Department of Physics, Okayama University, 
	Okayama 700-8530, Japan\\
	Telephone number: +81-86-251-7817\\ 
	Fax number: +81-86-251-7830\\
	E-mail : tanimoto@fphy.hep.okayama-u.ac.jp}
\address[sakuda]{High Energy Accelerator Research Organization (KEK),
	Tsukuba 305-0801, Japan}
\begin{abstract}
We set limits on the magnetic moment and charge radius of the tau
neutrino by examining an extra contribution to the electroweak process 
$e^+ e^- \rightarrow \nu \overline{\nu}\gamma$ using VENUS, TOPAZ and AMY 
results.
We find that $\kappa(\nu_{\tau}) < 9.1\times{10}^{-6}$
 (i.e.\,\,$\mu(\nu_{\tau}) < 9.1\times{10}^{-6}\mu_B,\,\mu_B = e/2m_e$) 
and $\langle r^2 \rangle < 3.1\times{10}^{-31}$cm$^2$
with Poisson statistics by combining their results.
Whereas, similar to this method, with the Unified Approach we find that
$\kappa(\nu_{\tau}) < 8.0\times{10}^{-6}$ and 
$\langle r^2 \rangle < 2.7\times{10}^{-31}$cm$^2$.
\end{abstract}
\end{frontmatter}
\clearpage
The electromagnetic properties of neutrinos have been vigorously
examined in recent years, since they are related
to the neutrino mass and to solar neutrino problems.
Neutrino oscillation between $\nu_{\mu}$ and $\nu_{\tau}$, which
means a finite neutrino mass, has become realistic since evidential
results by Super-Kamiokande~\cite{SK}.
Gninenko~\cite{Gninenko} has shown the bound on the tau neutrino
magnetic moment from the S-K atmospheric neutrino data to be
$1.3\times{10}^{-7}\mu_B$.
Chua and Hwang~\cite{Chua} recently remarked
that the third-generation neutrino magnetic moment induced
by leptoquarks might be of the order of ${10}^{-10} \sim {10}^{-13}\mu_B$.

If neutrinos have mass, the tau neutrino would be the most massive
among  $\nu_e$, $\nu_{\mu}$ and $\nu_{\tau}$.
Therefore, the tau neutrino magnetic moment might show a relatively large
value because the neutrino magnetic moment is estimated to be
proportional to its mass according to the standard model
extended to have the right-handed neutrino singlet ($\nu_R$),
$\mu_{\nu}=3eG_Fm_{\nu}/(8\pi^2\sqrt{2})=(3.20\times {10}^{-19})
m_{\nu}\mu_B$~\cite{fujikawa}.
Assuming $m_{\nu_{\tau}}<18.2$ MeV~\cite{Barate},
the upper limit on the magnetic moment 
is less than $5.8\times 10^{-12}\mu_B$,
where $e, G_F$, $m_{\nu}$ and $\mu_B$ are the electron charge
magnitude,
the Fermi coupling constant, the neutrino mass in eV and
the Bohr magneton ($=e/2m_{e}$), respectively.
Thereupon, we estimated the tau neutrino magnetic moment
and charge radius limits with
the $e^+e^-\rightarrow \nu \overline{\nu} \gamma$
results from three TRISTAN experimental groups.

We classify the methods to evaluate the tau neutrino magnetic moment
as follows:
(A)cosmological estimation, (B)fixed target experiments
($\nu_{\tau}e^-\rightarrow \nu_{\tau}e^-$) and
(C)$e^+e^-$ colliding beam experiments.
Method(A) gives a very strong upper limit of the magnetic moment such as
$\mu_{\nu_{\tau}}<6.2 \times {10}^{-11}\mu_B$~\cite{Elmfors},
$2\times {10}^{-12}\mu_B$~\cite{Raffelt} and
$6\times {10}^{-14}\mu_B$~\cite{Nussinov}.
However, these values are based on many cosmological assumptions.
Incidentally, Grifols and Mass\'o~\cite{Masso} have argued that
primordial nucleosynthesis also constrains the
neutrinos charge radii to satisfy
$\langle r^2 \rangle < 7 \times 10^{-33} {\rm cm}^2$.
Their argument, however, also has an implicit dependence
on the neutrino mass which may allow them to be evaded.
Method(B) gives $\mu_{\nu_{\tau}}< 5.4
\times {10}^{-7}\mu_B$~\cite{BEBC}.
It assumes the form factor ratio of $f_{D_{s}}/f_{\pi} = 2$
and $D_s$, $\overline{D_s}$ production cross section $=2.6 \mu {\rm b}$
to calculate $\nu_{\tau}$ flux,
because $\nu_{\tau}$ beam flux has to be produced and estimated by
$D_s$, $\overline{D_s}$ production.
Method(C) is the most direct.
Grotch and Robinett~\cite{Grotch} combined the results from
ASP, MAC and CELLO experiments well below the $Z^0$ resonance
and set the limits at the 90\% confidence level on the magnetic moment and
the charge radius of the tau neutrino,
$\mu_{\nu_{\tau}}<4\times {10}^{-6} \mu_B$ and
$\langle r^2 \rangle < 2\times{10}^{-31}{\rm cm}^2$,
respectively.
The other is from the experiments at the $Z^0$ resonance.
They give $\mu_{\nu_{\tau}}<4.4\times{10}^{-6}\mu_B$~\cite{Abreu}
and $3.3\times {10}^{-6}\mu_B$~\cite{L3}.

At energies well below $Z^0$,
the dominant contribution to the process
$e^+e^-\rightarrow\nu\overline{\nu}\gamma$
involves the exchange of a virtual photon~\cite{Grotch}.
The dependence on the magnetic moment comes from its direct coupling
to the virtual photon, and the observed photon is the result of
the initial-state Bremsstrahlung.

While the results of the TRISTAN experiments (VENUS, 
TOPAZ and AMY ~\cite{VenusTopazAmy} collaborations) have been used
to set limits on supersymmetric particles,
we will make use of them here to set limits on the tau neutrino magnetic
moment and charge radius.

\begin{figure}[htbp]
\begin{center}
	\includegraphics{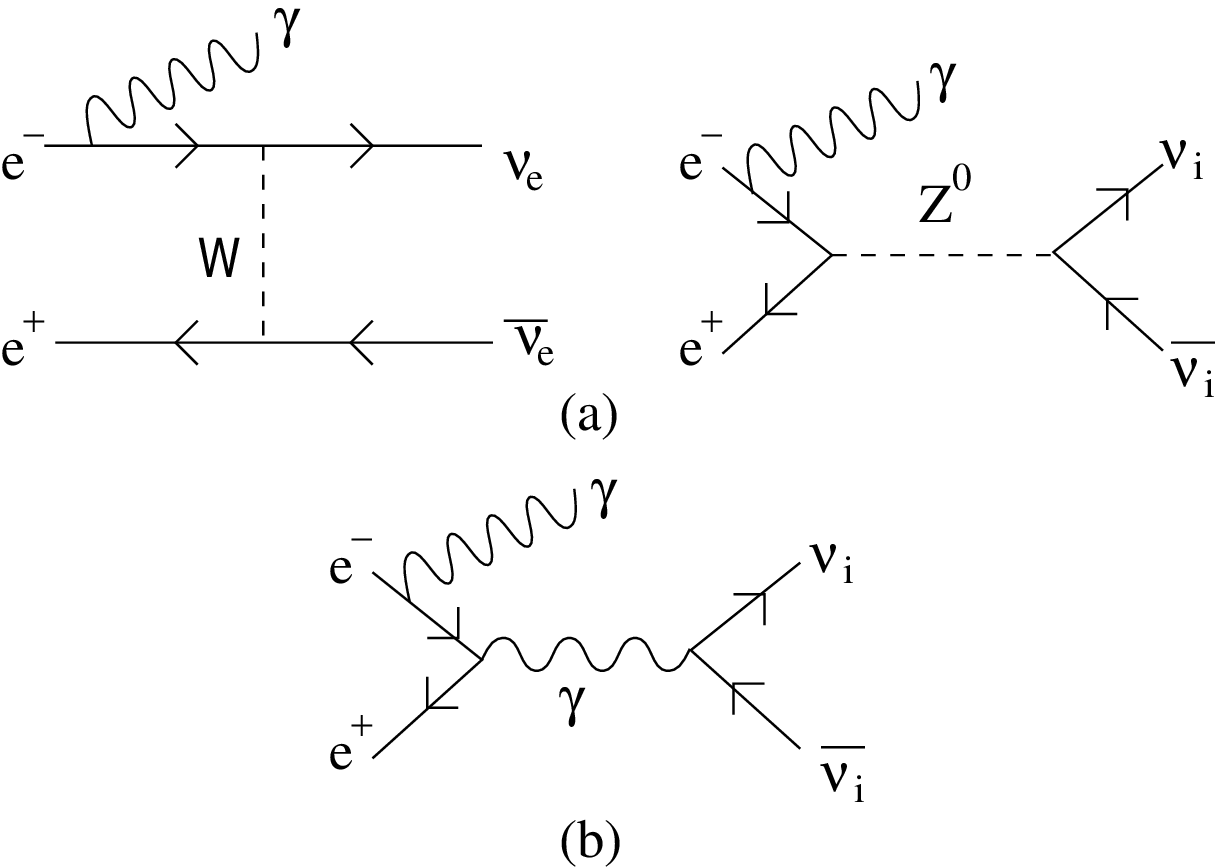}
	\caption{Diagrams leading to the process 
	$e^+e^- \rightarrow \nu_i \overline{\nu_i} \gamma$ 
	due to (a) standard model processes and (b) 
	contributions from anomalous neutrino electromagnetic coupling.}
	\label{fig1}
\end{center}
\end{figure}

The standard expression~\cite{Gaemers} for the cross section
for the process $e^+e^- \rightarrow \nu \overline{\nu} \gamma$
due to $Z^0$ and $W$ exchange (Fig.\ref{fig1}(a)) is
\begin{eqnarray}
\label{eq:crosec_wz}
\frac{d\sigma}{dx\,dy}
&=&
\frac{{G_F}^2 \alpha}{6\pi^2}\nonumber \\
& & \cdot
\left\{
 	\frac{M^4_Z
	    \left\{ N_{\nu}(g_V^2 + g_A^2) + 2(g_V + g_A)[1 - s(1-x) / M_Z^2]
	    \right\} }{[s(1-x) - M_Z^2]^2 + (M_Z {\Gamma}_Z)^2}
+2 \right\}   \nonumber \\
& & \cdot
\frac{s}{x(1 - y^2)}
\left[(1 - x)(1 - x/2)^2 + x^2 (1 - x) \frac{y^2}{4} \right],
\end{eqnarray}
where $x=E_{\gamma}/E=2E_{\gamma}/\sqrt{s}$ is the photon energy
in units of the incident beam energy,
$y=\cos\theta_\gamma$ is the direction cosine of the photon momentum
with respect to the incident beam direction,
$\alpha$ is the fine-structure constant,
$s$ is the square of the center of mass energy,
$N_{\nu}$ is the number of low-mass neutrino generations,
$M_{Z}$ is the mass of the $Z^0$,
$\Gamma_{Z}$ is the total width of $Z^0$,
$g_V = - 1/2 + 2 \sin^2\theta_W$($\theta_W$ is the weak mixing angle)
and $g_A = -1/2$.
It is worth noting that the $(g_V^2 + g_A^2)$ term of equation
(\ref{eq:crosec_wz}) arises from the square of the s-channel $Z^0$ amplitude,
the `2' term from the square of the t-channel $W$-exchange amplitude,
and the $(g_V + g_A)$ term from $Z^0-W$ interference.
We now allow for a neutrino electromagnetic interaction given by
the vertex $-ie({\gamma_{\mu}}F_1(q^2)+(\kappa /2m_e)
{\sigma}_{\mu \rho}q^{\rho})$,
where we express $F_1(q^2)$ as $q^2 \langle r^2 
\rangle /6 $ in order to extract a limit on a possible charge
radius\footnote{We consider here only Dirac neutrinos
since Majorana neutrinos are well known to have quite
different electromagnetic properties,
in particular, they cannot possess a magnetic moment.}.
We will include such a contribution only for the tau neutrino because
the limits which were already obtained for $\nu_e$ and $\nu_{\mu}$
are more stringent than the limit we will be obtaining for the tau neutrino.
We obtain the additional contributions from the diagram
of Fig.\ref{fig1}(b) to the cross section,
\begin{eqnarray}
\label{eq:crosec_vg}
\frac{d\sigma}{dx\,dy}
& = &
\frac{\alpha^3}{3}
\Biggl\{ \frac{ 2 {\langle r^2 \rangle}^2 s(1-x)}{9}
	+ \frac{\kappa^2}{m_e^2} \nonumber \\
& &     - \frac{g_V \langle r^2 \rangle M_Z^2 s(1-x) (1-s(1-x)/M_Z^2)}
	{3 \sin^2{\theta}_W \cos^2{\theta}_W[s(1-x) - M_Z^2]^2
	+(M_Z \Gamma_Z)^2}\Biggr\}\nonumber \\
& & \cdot \frac{[(1-x/2)^2 + x^2 y^2 /4]}{x(1-y^2)}.
\end{eqnarray}
We have integrated (\ref{eq:crosec_vg}) over the relevant range given
in Table~\ref{table:integRange} for each experiment~\cite{VenusTopazAmy}.
In Table~\ref{table:integRange}, $x_T(=E_{T\gamma}/E)$ is
the photon transverse energy normalized to the beam energy,
and $\epsilon$ is the overall efficiency for each data sample.
For instance, integrating (\ref{eq:crosec_vg}) over the VENUS
kinematical region, we changed the variable from $x$ to $x_T$ with Jacobian,
then integrated it over the region $0.13 < x_T < 1$ and
$\cos130.3^{\circ}\leq y\leq\cos50.0^{\circ}$.
We also applied a similar method to the other experiments.
Table~\ref{table:singlePhoton} is a summary of
the number of single-photon candidates for each experimental result.
It is worth noting that there is no interference
between (\ref{eq:crosec_wz}) and (\ref{eq:crosec_vg}),
since the anomalous contribution given in equation (\ref{eq:crosec_vg})
flips helicity, but the standard model contribution given in 
equation (\ref{eq:crosec_wz}) does not~\cite{Deshpande}.

We obtained the upper limits on the number of signal events
for the observed events and the expected background using two methods:
One is Poisson statistics~\cite{PDG},
\begin{equation}
        1-\alpha = 1-
        \frac{e^{-(n_B+N)}\displaystyle\sum_{n=0}^{n_0}
        \frac{(n_B+N)^n}{n!}}%
        {e^{-n_B}\displaystyle\sum_{n=0}^{n_0}\frac{n_B^n}{n!}} ,
        \label{eq:poissonWithBackgrounds}
\end{equation}
where $n_0$ is the number of the single-photon candidates
which each experiment has obtained,
$n_B$ is the mean for the sum of all backgrounds
and $N$ is the desired upper limit on the unknown mean for the signal
with confidence coefficient $\alpha$.
The other is the Unified Approach~\cite{Feldman}.
We applied $n_B = n_0$ to both methods because
each $n_0$ in Table~\ref{table:singlePhoton} could be explained
by the sum of the number of physically expected events and that of
non-physical backgrounds.
Then, we required
\begin{equation}
N_{i} > \sigma_{i}\times \epsilon_{i}\times \int L_{i}\,dt
\end{equation}
for each experiment, where $N_{i}$ is the upper limit at 90\% C.L.
on the number of signal events, $\sigma_{i}$
is the cross section obtained from integration of (\ref{eq:crosec_vg})
over the relevant ranges, $\epsilon_{i}$ is the overall efficiency,
and $\int L_{i}\,dt$ is the integrated
luminosity for each experiment, i.e., $i$ means VENUS, TOPAZ, and so on.
Table~\ref{table:singlePhoton} contains the upper limits of
the tau neutrino magnetic moment for each experiment.

We combined the bounds on the magnetic moment at the 90\% C.L.
at TRISTAN using Poisson statistics,
\begin{equation}
	\kappa < 9.1 \times {10}^{-6},
	\label{eq:Pkappa}
\end{equation}
and also derived the bounds on the charge radius of the tau neutrino
for the experiments (Table~\ref{table:singlePhoton})
and combined them at the 90\% C.L.,
\begin{equation}
	\langle r^2 \rangle < 3.1 \times 10^{-31} {\rm cm^2}.
\end{equation}

In addition, using the Unified Approach,
we obtained the combined bound on the magnetic moment
at the 90\% C.L.,
\begin{equation}
	\kappa < 8.0 \times {10}^{-6}
	\label{eq:UAkappa}
\end{equation}
and charge radius
\begin{equation}
	\langle r^2 \rangle < 2.7 \times 10^{-31} {\rm cm^2}.
	\label{eq:UAmag}
\end{equation}
The obtained results (\ref{eq:Pkappa})-(\ref{eq:UAmag}) give
upper limits comparable to those obtained from other $e^+ e^-$
colliding beam experiments.

We have reported on the bound on the tau neutrino magnetic moment
and charge radius from the TRISTAN experiments
and have obtained the bound from single photon production
cross section at TRISTAN,
$9.1 \times {10}^{-6}\mu_B$
and
$3.1\times 10^{-31} {\rm cm^2}$ at 90\% C.L.
using Poisson statistics, and
$8.0\times {10}^{-6} \mu_B$ and
$2.7\times 10^{-31} {\rm cm^2}$ at 90\% C.L.
using Unified Approach.
They are still far above what is predicted by the standard
electroweak theory extended to include massive neutrinos
although comparable to the results from other $e^+ e^-$ colliding
experiments.

\begin{ack}
We would like to thank G.J. Feldman and R.D. Cousins for
providing the calculations for the upper end of the signal mean
using Unified Approach.
\end{ack}

\clearpage
\clearpage
\begin{table}[htbp]
\caption{Summary of the criteria for TRISTAN experiments. 
$\epsilon$ is the overall efficiency for each data sample.}
\label{table:integRange}
\begin{tabular}{rlcccc}\hline
      &$\sqrt{s}$[GeV]& $x_T,\,x$  & $y=\cos\thega$                           & $E_{\gamma}$[GeV] & $\epsilon$ \\ \hline
VENUS & 58            & $x_T>0.13$ & $50.0^{\circ}\leq\thega\leq130.3^{\circ}$&                   & 0.57\\
TOPAZ & 58            & $x_T\geq 0.12$ & $|\cos\thega|\leq 0.8$               & $\geq 4.0$        & 0.27\\
AMY 1 & 57.8          & $x>0.175$ & $|\cos\thega|<0.7$                        &                   & 0.44\\
    2 & 57.8          & $x>0.175$ & $|\cos\thega|<0.7$                        &                   & 0.64\\
    3 & 57.8          & $x>0.125$ & $|\cos\thega|<0.7$                        &                   & 0.58\\ 
    4 & 57.8          & $x>0.125$ & $|\cos\thega|<0.7$                        &                   & 0.57\\  \hline
\end{tabular}
\end{table}
\clearpage
\begin{table}[htbp]
\caption{Summary of the upper limits of the tau neutrino magnetic moment 
and the charge radius for each experimental result with Poisson 
statistics and Unified Approach, and their combined results.
$n_0$ is the number of the single-photon candidates which 
each experiment has obtained. Numbers in parentheses indicate the 
expected number of events originating from $W$ and $Z^0$ exchange.}
\label{table:singlePhoton}
\begin{center}
\begin{tabular}{@{}rcl@{}lcccc}\hline
 & & & &\multicolumn{2}{l}{Poisson statistics}&\multicolumn{2}{l}{Unified Approach}\\
 & $\int L\,dt$ & $n_0$ & & $\kappa $ & $\langle r^2 \rangle $
 & $\kappa $ & $\langle r^2 \rangle $\\ 
 & [pb$^{-1}$] & & & [$\times{10}^{-6}$] & [$\times{10}^{-31}{\rm cm}^2$] & [$\times{10}^{-6}$]
 & [$\times{10}^{-31}{\rm cm}^2$]\\ \hline
VENUS & 164.1 & 8&($3.9^{+4.2}_{-2.8}$)    & $<$ 13.  & $<$ 4.6 & $<$ 13. & $<$ 4.6\\ 
TOPAZ & 213   & 5&(3.1)    & $<$ 13. & $<$ 4.6 & $<$ 13. & $<$ 4.5\\ 
\begin{tabular}{@{}r@{}r@{}r@{}r}
AMY 1 \\
    2 \\
    3 \\
    4
\end{tabular} & 
\begin{tabular}{@{}r@{}r@{}r@{}r}
55 \\
91 \\
56 \\
99
\end{tabular} &  
\begin{tabular}{@{}r@{}r@{}r@{}r}
0 \\
2 \\
2 \\
2
\end{tabular} &
\multicolumn{1}{l@{}}{$\left.\makebox{\rule[-6ex]{0pt}{6ex}}\right\}$(7.2)} &
\begin{tabular}{@{}c@{}c@{}c@{}c}
$<$ 16.\\
$<$ 14.\\
$<$ 16.\\
$<$ 12.
\end{tabular} & 
\begin{tabular}{@{}c@{}c@{}c@{}c}
$<$ 5.9\\
$<$ 4.9\\
$<$ 5.6\\
$<$ 4.2
\end{tabular} & 
\begin{tabular}{@{}c@{}c@{}c@{}c}
$<$ 17.\\
$<$ 14. \\
$<$ 16.\\
$<$ 12.
\end{tabular} & 
\begin{tabular}{@{}c@{}c@{}c@{}c}
$<$ 6.0 \\
$<$ 4.9\\
$<$ 5.6\\
$<$ 4.2
\end{tabular}\\ \hline\hline
\multicolumn{4}{c}{Combined results} & $<$ 9.1 & $<$ 3.1 & $<$ 8.0 & $<$ 2.7\\\hline
\end{tabular}
\end{center}
\end{table}

\clearpage
\begin{thebibliography}{99}
\bibitem{SK}
Super-Kamiokande Collab., Y. Fukuda et al., Phys. Rev. Lett. 81 (1998) 1562.
\bibitem{Gninenko}
S.N. Gninenko, Phys. Lett. B 452 (1999) 414.
\bibitem{Chua}
C.-K. Chua and W.-Y. P. Hwang, Phys. Rev. D 60 (1999) 073002.
\bibitem{fujikawa}
B.W. Lee and R.E. Shrock, Phys. Rev. D 16 (1977) 1444;
K. Fujikawa and R.E. Shrock, Phys. Rev. Lett. 45 (1980) 963.
\bibitem{Barate}
ALEPH Collab., R. Barate et al., Eur. Phys. J. C 4 (1998) 433.
\bibitem{Elmfors}
P. Elmfors, K. Enqvist, G. Raffelt and G. Sigl, Nucl. Phys. B 503 (1997) 3.
\bibitem{Raffelt}
G. Raffelt, Phys. Rev. Lett. 64 (1990) 2856.
\bibitem{Nussinov}
S. Nussinov and Y. Rephaeli, Phys. Rev. D 36 (1987) 2278.
\bibitem{Masso}
J.A. Grifols and E. Mass\'o, Mod. Phys. Lett. A 2 (1987) 205.
\bibitem{BEBC}
A.M. Cooper-Sarkar et al., Phys. Lett. B 280 (1992) 153.
\bibitem{Grotch}
H. Grotch and R.W. Robinett,  Z. Phys. C 39 (1988) 553.
\bibitem{Abreu}
DELPHI Collab., P. Abreu et al., Z. Phys. C 74 (1997) 577.
\bibitem{L3}
L3 Collab., M. Acciarri et al., Phys. Lett. B 412 (1997) 201.
\bibitem{VenusTopazAmy}
VENUS Collab., N. Hosoda et al., Phys. Lett. B 331 (1994) 211;
TOPAZ Collab., T. Abe et al., Phys. Lett. B 361 (1995) 199;
AMY Collab., Y. Sugimoto et al., Phys. Lett. B 369 (1996) 86.
\bibitem{Gaemers}
K.J.F. Gaemers, R. Gastmans and F.M. Renard, Phys. Rev. D 19 (1979) 1605.
\bibitem{Deshpande}
N.G. Deshpande and K.V.L. Sarma, Phys. Rev. D 43 (1991) 943.
\bibitem{PDG}
Review of Particle Properties, Phys. Rev. D 54 (1996) 1. 
\bibitem{Feldman}
G.J. Feldman and R.D. Cousins, Phys. Rev. D 57 (1998) 3873.
\end{thebibliography}
\end{document}